\documentclass[12pt]{iopart}

\usepackage[dvips]{graphicx}

\usepackage{iopams}  
\usepackage{color,epsfig}  
\begin{document}

\title{Open Problems in $\alpha$ Particle Condensation}
\author{Y.~Funaki$^1$, M.~Girod$^2$, H.~Horiuchi$^{3,4}$, G.~R\"opke$^5$, P.~Schuck$^{6,7,8}$, A.~Tohsaki$^3$, and T.~Yamada$^9$}
\address{$^1$Institute of Physics, University of Tsukuba, Tsukuba 305-8571, Japan}
\address{$^2$CEA/DAM/DIF, F-91297 Arpajon, France}
\address{$^3$Research Center for Nuclear Physics (RCNP), Osaka University, Osaka 567-0047, Japan}
\address{$^4$International Institute for Advanced Studies, Kizugawa 619-0225, Japan}
\address{$^5$Institut f\"ur Physik, Universit\"at Rostock, D-18051 Rostock, Germany}
\address{$^6$Institut de Physique Nucl\'eaire, CNRS, UMR 8608, Orsay, F-91406, France}
\address{$^7$Universit\'e Paris-Sud, Orsay, F-91505, France} 
\address{$^8$Laboratoire de Physique et Mod\'elisation des Milieux Condens\'es, CNRS et Universit\'e Joseph Fourier, 25 Av.~des Martyrs, BP 166, F-38042 Grenoble Cedex 9, France}
\address{$^9$Laboratory of Physics, Kanto Gakuin University, Yokohama 236-8501, Japan}


\begin{abstract}
$\alpha$ particle condensation is a novel state in nuclear systems.
We briefly review the present status on the study of $\alpha$ particle condensation and
 address the open problems in this research field:~$\alpha$ particle condensation in heavier systems
 other than the Hoyle state, linear chain and $\alpha$ particle rings, Hoyle-analogue states
 with extra neutrons, $\alpha$ particle condensation related to astrophysics, etc.
\end{abstract}


\section{Introduction}

Clustering in strongly interacting Fermi systems is a very important issue in many body physics. 
Most well known is two body clustering, i.e. pairing, since it gives rise to superconductivity and superfluidity. 
Much less studied is the phenomenon of formation of heavier clusters, like trions and quartets. 
These heavier clusters have so far mostly been relevant and studied in nuclear systems. 
This stems from the fact that in nuclear physics there are more than two species of fermions,
 i.e.~four different nucleons, which are the proton and neutron, both spin up or down. 
They all attract one another forming a strongly bound quartet, i.e.~the $\alpha$ particle which is
 the smallest closed shell nucleus with a bosonic nature. 
It is known that $\alpha$-clustering aspects as well as mean-filed aspects are essential features
 to understand the structure of light nuclei.

Recently it has been pointed out that certain states in self conjugate nuclei around the $\alpha$ particle disintegration
 threshold can be described as product states of $\alpha$ particles, all in the lowest $0S$ state.
We define a state of condensed $\alpha$ particles in nuclei as a bosonic product state in good approximation,
 in which all bosons occupy the lowest quantum state of the corresponding bosonic mean field potential.
A typical example is the Hoyle state, $0^{+}_{2}$ state at $E_x = 7.65$ MeV in $^{12}$C,
 the wave function of which is described by a product state of three $\alpha$'s with $70$~\% probability.
In heavier $A=4n$ nuclei Hoyle-like states are predicted to exist in low density states,
 close to the $n\alpha$ disintegration threshold.
Studies on $\alpha$ particle condensation are quite advanced~\cite{ref:2}, 
 but still many open questions and problems exist. 
We want to address a couple of them in this paper.
 

\section{$\alpha$ particle condensation in heavier systems, other than the Hoyle state}

Theory predicts that analogues to the Hoyle state in $^{12}$C with its three $\alpha$ particles condensed into a common $0S$ orbit also should exist around the n$\alpha$ threshold in heavier self conjugate nuclei~\cite{ref:2,thsr,tohsaki_nara,yamada04}. We have made a prediction that the 6-th $0^+$ state at $15.1$ MeV in $^{16}$O should be such a state~\cite{funaki08}. Even without any calculations, one realizes its strong similarity with the Hoyle state:
\noindent

i)~It is strongly excited by inelastic electron scattering and, thus, its monopole transition matrix element is quite large~\cite{monopole}, $M_{\rm calc} =1.0$ fm$^2$ $(^{12}{\rm C}(0_2^+)$: $M=5.4$ fm$^2$).

\noindent

ii)~It is situated just a couple of hundred keV above threshold: Hoyle in $^{12}$C at $380$ keV above threshold (7.27 MeV), 6-th $0^+$ state $700$ keV above 4 $\alpha$ particle threshold (14.4 MeV).

\noindent

iii)~Though the width of the 15.1 MeV state in $^{16}$O (160 keV) is much larger than the one of the Hoyle state (8.5 eV), it still is unusually small for a nuclear state so high up in energy. 

The width depends crucially on its position under the Coulomb barrier. The 15.1 MeV state in $^{16}$O is about twice as much above threshold as the Hoyle state. Barrier height also may be lower in $^{16}$O because of stronger Coulomb repulsion of four vs three $\alpha$ particles. More decay channels are open because of higher energy. Still, they are not so many because shell model type of states have very little overlap with $\alpha$ condensate states. Though those features easily can explain the larger width in $^{16}$O, it is still surprisingly small. Unfortunately, measurements of the 15.1 MeV state in $^{16}$O are much less advanced than for the Hoyle state. Inelastic electromagnetic form factors are very much in need. The Hoyle state has very clearly been identified from a three $\alpha$ coincidence measurement~\cite{ref:3}, hinting to its $\alpha$-gas like structure. A 4$\alpha$ coincidence measurement in $^{16}$O with determination of the invariant mass, would be extremely useful. Hope comes from the analysis of an experiment at LNS Catania with the CHIMERA detector with 4$\alpha$ particle coincidences~\cite{ref:4_borderie}. However, we need more experiments of this kind. Also for more $\alpha$'s. There are probably many states in $^{16}$O which can be interpreted as excited states out of the condensate. A conspicuous rotational band at around 17 MeV with $I = 0^+$, $2^+$, $4^+$, and $6^+$ with a very large moment of inertia is well known since several decades~\cite{ref:5_chevalier}. It also has recently been reanalyzed by Freer et al.~\cite{ref:6_freer}.
It is of great importance to study the cluster structures in the 15 - 18 MeV range in greater detail.

One may reasonably think that also $^{20}$Ne, $^{24}$Mg possess $\alpha$ gas states. The Ikeda threshold~\cite{ikeda68} increases fast though: 19.17 MeV for $^{20}$Ne, 28.48 MeV for $^{24}$Mg, 38.46 MeV for $^{28}$Si, etc. If the experimenters were able to excite those nuclei into a large amplitude breathing mode, at some point surely they would cluster into a coherent state of $\alpha$ particles. Eventually for $n=8$ - $10$ $\alpha$'s there will not exist a Coulomb barrier any longer and the condensate gas will start to expand coherently as a soft Coulomb explosion mode. There exist proposals to excite those large amplitude $0^+$ states by Coulomb excitation with projectiles of relativistic energies~\cite{ref:7_tanihata}. Some events related with the Coulomb explosion are known in light nuclei since long from emulsion tracks~\cite{ref:9}. A dream may be to dissociate $^{40}$Ca into 10 $\alpha$'s. However, the $n\alpha$ nuclei need not to disintegrate totally into $n\alpha$'s. There may subsist a non disintegrated core, as e.g. ${^{32}{\rm S}^\ast}= {^{16}{\rm O}} + 4\alpha$, ${^{52}{\rm Fe}^\ast} = {^{40}{\rm Ca}} +3\alpha$, etc. Such situations have been discussed by von Oertzen~\cite{ref:10_oertzen}, Brenner~\cite{ref:11_brenner}, and Ogloblin~\cite{ref:12_ogloblin}. Still a clear finger print of such $\alpha$ particle condensates on top of an inert core must be experimentally verified. However, in $\alpha$ decay processes a bosonic enhancement seems having been seen~\cite{ref:2}.

\section{Linear chain states and $\alpha$ particle rings}

$\alpha$ particle cluster physics excited nuclear physicist's imagination from the very early days when Morinaga 
interpreted the Hoyle state as a linear chain of three $\alpha$ particles~\cite{morinaga56}. The linear chain structure of alpha clusters 
has been discussed by many people theoretically and experimentally~\cite{ref:14}. Famous examples are the ones of 
$^{12}$C, $^{16}$O, and $^{24}$Mg.  However, the states in $^{12}$C and in $^{24}$Mg which were assigned to have linear 
alpha-chain structures are now regarded to be inappropriate.  For example, in $^{12}$C, the Hoyle state which is the 
second $0^+$ state at 7.65 MeV, was assigned to have $3\alpha$ linear chain structure but it is now considered to have a 
$3\alpha$-gas-like structure.  However, this does not mean that the existence of linear alpha-chain structures is 
impossible. In fact, in the case of $^{12}$C, there are some reports that the third $0^+$ state around 10 MeV has 
$3\alpha$ chain structure, although it is slightly bent away from the linear shape~\cite{ref:15_feldmeier,ref:16_enyo}

For the case of $^{16}$O, as mentioned, Chevalier et al.~\cite{ref:5_chevalier} very early found a rotational band 
 around 17 MeV in $^{16}$O with a huge moment of inertia which they said can only be understood if the corresponding 
 state is interpreted as a linear chain of four $\alpha$ particles.  
While many calculations have supported the idea of $4\alpha$ linear chain, there are some discussions
 with the $4\alpha$ OCM (Orthogonality Condition Model)~\cite{funaki08,funaki_yamada09} and potential model~\cite{ref:17_ohkubo_to_be_published} 
 which try to explain the rotational states of Chevalier et al.~by relating them to the $4\alpha$ condensate state.  
About the existence of chains of six $\alpha$ particles there is a rather recent discussion~\cite{ref:18_chappell}.

Let us dwell a little further on these hypothetical linear chain states. The usual picture of the linear alpha-chain 
structure is expressed by the use of Brink wave function which has alpha clusters located on a line with equal 
inter-cluster distances. This means that Brink's description assumes a quasi rigid body picture of the chain state. 
One can improve the wave function by introducing small-amplitude oscillations of alpha clusters around their equilibrium 
positions. We now have some doubt about this common picture of linear alpha-chain structure on the basis of our 
experience with the investigation of alpha-condensate-like states expressed by the THSR (Tohsaki-Horiuchi-Schuck-R\"opke)
wave function~\cite{thsr}. Our studies of 
alpha-condensate-like states teach us that, in three or many alpha systems, when alpha clusters are clearly separated, 
they do not necessarily keep any fixed geometrical arrangement but they move freely in the whole nuclear volume like 
a gas. From this knowledge, we can make the conjecture that the alpha clusters arranged on a line may move rather 
freely along the line instead of staying around some equilibrium points. We checked the validity of this conjecture 
in the case of three alpha clusters on a line in the following way. We calculated the energy obtained by the Brink 
wave function with three alpha clusters arranged on a line with equal distance $d$ between neighboring clusters. Let 
us denote the minimum energy with respect to $d$ as $E({\rm Brink})$. Next we calculated the energy obtained by a 
one-dimensional THSR wave function having the form,

\begin{equation}
{\cal A} [\exp\{ -\sum_i ( (X_i^2 + Y_i^2)/(b^2/2) + Z_i^2/(B^2/2) )\} 
(\phi(\alpha))^3 ] . 
\end{equation}

Let us denote the corresponding minimum energy with respect to $B$ as $E(1d-{\rm THSR})$. We found that 
$E(1d-{\rm THSR})$ is lower than $E({\rm Brink})$ by about 3.5 MeV. This  result supports our conjecture that even 
in a linear configuration the $\alpha$'s can preferably be in a condensate. In condensed matter such linear Bose 
condensates are called Tonks-Girardeau gases having quite particular properties~\cite{ref:19,pitaevskii03}. This stems from the 
fact that the ideal compact bosons cannot get around one another and, therefore, behave as hard core bosons. The fact 
that two bosons cannot be on the same spot turns them effectively into fermions. 
It would be interesting to see to which extent such a feature also is born out in those linear chain condensates.
It is well known that the $\alpha$-$\alpha$ interaction has a rather strong short-range repulsion
 if the relative motion energy is not so high.  
Thus we need to investigate whether this short-range repulsion remains active
 in the linear chain configuration of three or more $\alpha$'s.

We think that the alpha-condensate-like structure gives us new physics, even for the problem of linear chain structures 
of alpha clusters and we need further investigations in this direction. The linear chain structures probably are 
strongly stabilized if a couple of neutrons are added.  Actually the linear chain has recently be considered by 
Itagaki et al.~in $^{16}$C~\cite{ref:20_itagaki}.  They concluded that there exists a specific molecular orbit 
configuration of four excess neutrons for which linear chain structure is stabilized especially for bending mode.  
In such a case, it would be extremely interesting whether the fact that extra neutrons are between three 
$\alpha$'s hinders  condensation to be born out to some extent or even totally, see also discussion in Sec.~\ref{sec:4}.

\begin{figure}[htbp]
\begin{center}
\includegraphics[width=140mm,height=80mm]{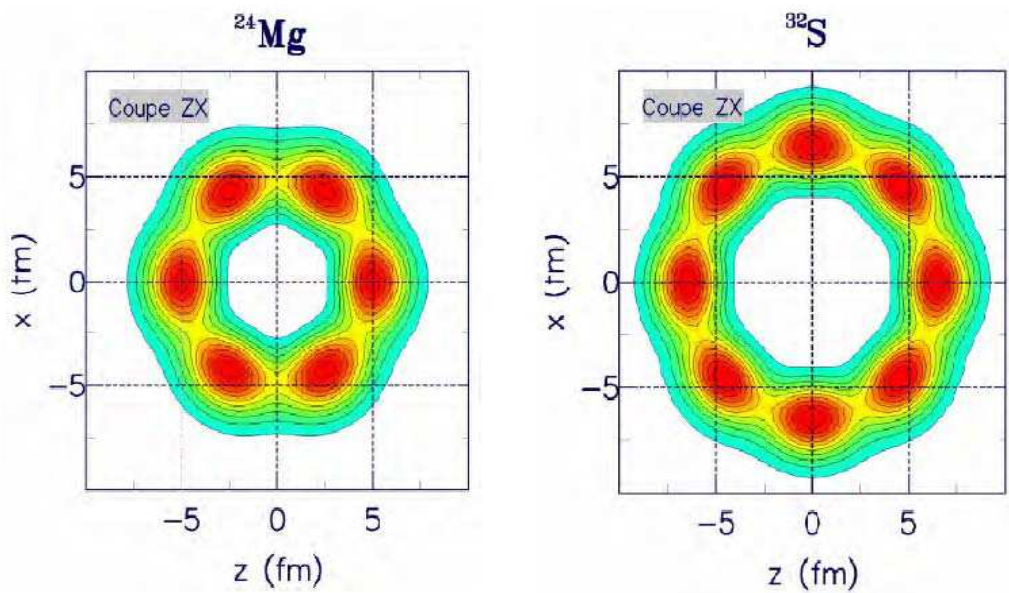}
\caption{Left (right) panel shows the $6\alpha$ ($8\alpha$) ring structure in $^{24}$Mg ($^{32}$S)
with constrained HFB calculations~\cite{girod}.}
\label{fig:rings_32S}
\end{center}
\end{figure}

But what about rings instead of chains? 
Such configurations also have been discussed very early in the 
 literature~\cite{ref:21}. 
We here want to show a very preliminary study of a constrained HFB calculation with the 
 D1S force~\cite{girod}. 
The constraint was to make the nucleus more and more oblate but the code allowed for 
 spontaneous breaking of rotational symmetry. 
At some point of low density the system preferred to break up into a 
 ring of let's say eight $\alpha$ particles for $^{32}$S. 
But a ring with six $\alpha$'s also has been found 
 in the same way. 
The raw result is shown in Fig.~\ref{fig:rings_32S}. 
Of course, the energies of these states which come from a pure mean 
 field HFB calculation are much too high. 
However, the clustering probably is real. 
We have similar experience from the Hoyle state. 
Taking out the spurious center-of-mass (c.o.m.) motion of the individual $\alpha$ particles may bring down the energy to 
 more realistic values.  
But their existence remains pure speculation for the moment. 
Also the rings may  considerably be stabilized by adding two neutrons per link. 
In the case of six $\alpha$'s twelve neutrons may give the most 
 binding. 
Whether chains or rings are more stable is an open problem which, obviously, may depend on 
 the number of $\alpha$'s involved.
We see that the chapter of $\alpha$ chains or even rings is by far not closed 
and many things and surprises can be expected in the future.



\section{Hoyle-type states with extra neutrons}\label{sec:4}

Besides the $4n$ nuclei, one can expect cluster-gas states composed of alpha clusters with extra neutrons (as well as deuterons and tritons etc.) around their cluster disintegrated thresholds in $A \not= 4n$ nuclei,  in which all clusters are in their respective $0S$-orbits, similar to the Hoyle state with the $(0S_{\alpha})^3$ configuration. These states, thus, can be called ``Hoyle-analogues'' in non-self-conjugated nuclei. It is an intriguing subject to investigate whether or not  Hoyle-analogue states exist in $A \not= 4n$ nuclei, for example for the simplest cases, $^{13}$C ($^{11}$B), composed of $3\alpha$'s and one valence neutron ($2\alpha$'s and one $t$). The pertinent question is in which way the extra neutrons will influence the Hoyle state. 

\begin{figure}[t]
\begin{center}
\includegraphics[width=0.3\hsize]{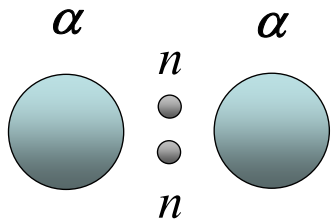}
\caption{Schematic figure of $2\alpha+2n$ structure in $^{10}$Be.}
\label{fig:10Be}
\end{center}
\end{figure}

Let us consider the case of $^{10}$Be when the two neutrons are in a $\pi$ orbit, see Fig.~\ref{fig:10Be}. For example is a Brink type wave function more adequate for a gas-like state than a THSR wave function? Or, on the contrary, can we expect that such novel Hoyle-like states have in addition to the $\alpha$ particles also the extra neutrons in loosely bound $0S$-orbits, not influencing very much the condensates? The answer to these questions probably very much depends on how the extra neutrons arrange themselves around the $\alpha$ particles. Very likely, the most easy cases are the ones where one just has one extra neutron, like in $^{9}$Be or $^{13}$C. The extra neutron in $^{9}$Be gives about 1.5 MeV extra binding. An OCM calculation has recently been performed by Yamada et al.~\cite{yamada08} for $^{13}$C. The resulting picture concerning the Hoyle-like states is the following. In the case of the valence neutron mainly in the $p$-orbit, the attractive effect of the $p$-wave $\alpha$-$n$ interaction induces 1)~a reduction of the nuclear radius of the Hoyle state and 2)~a strong coupling of the $3\alpha$ motion with the valence neutron, namely,  the coupling of the cluster configurations such as $^{12}$C($0^+,2^+$)+$n$ and $^9$Be($3/2^{-},1/2^{-})$+$\alpha$ etc. Consequently, the $3\alpha$ condensate aspect in $1/2^{-}$ states of $^{13}$C is significantly deteriorated, although the $1/2^{-}_{4}$ state around the $3\alpha$+$n$ threshold, corresponding to the 12.4-MeV state recently observed by Kawabata et al.~\cite{sasamoto06},  has a somewhat large component of the Hoyle state in comparison with the other states. 

On the other hand, in the case of the valence neutrons mainly in the $s$-orbit, the situation is very different from the case of the neutron in the $p$-orbit, due to the weak attractive effect of the $s$-wave $\alpha$-$n$ interaction. It should be recalled that the $p$-wave $\alpha-n$ interaction produces the resonant states of $3/2^-$ and $1/2^-$ in $^5$He, while the $s$-wave does not. The results of the OCM calculation show that the $1/2^+_3$ state (corresponding to the 12.1-MeV state) around the $3\alpha+n$ threshold has a dilute condensate character, in which the three alpha clusters occupy an identical $0S$ orbit with $55~\%$ probability, indicating the state being a candidate of the Hoyle-like state with the $[(0S_\alpha)^3(s_\nu)]$ configuration.

The spin-orbit splitting in $^{13}$C, $3/2^-$-$1/2^-$, is also an interesting quantity to extract the information on the structure of the nuclear core, because the one-body spin-orbit potential for the extra neutron depends on the density distribution of the nuclear core. Since the Hoyle state has about triple volume with respect to the ground state, it can be expected that the spin-orbit splitting felt by an extra neutron is correspondingly weakened as compared with the one of the ground-state configuration~\cite{yamada06}. Thus, the spin-orbit splitting is one of the good tools to search for the Hoyle state coupled with an extra neutron.

Concerning the case of $^{11}$B, a Hoyle-analogue state with the main configuration of $[(0S)^2_{\alpha}(0S)_t]$ was explored with an OCM calculation~\cite{yamada09}. It was indicated that the $12.56$-MeV state ($J^\pi=1/2^+$) around the $\alpha+\alpha+t$ threshold observed in the $^7$Li($^{7}$Li,$^{11}$B$^*$)$t$ reaction etc.~\cite{soic04,curtis05,charity08} is a candidate for the Hoyle-analogue state from the analyses of the single-cluster motions in $^{11}$B.

Generally speaking one can expect that adding a lot of neutrons to Hoyle like states will hinder their formation because of the Pauli principle. In chains and rings, as discussed before, neutron pairs between the $\alpha$ particle links may substantially stabilize those configurations. However, it is not clear whether these structures still show $\alpha$ condensation aspects.

\section{$\alpha$ particle condensation related to astrophysics}\label{sub:5}

The cluster composition of compact stars is another field where eventually condensation of $\alpha$ particles may play a role. Recent calculations~\cite{lattimer_svesty,toki-shen,shimiyoshi-roepke} investigate the $\alpha$ particle content of infinite nuclear matter as a function of density, asymmetry, and temperature. Non-negligible $\alpha$ particle mass fractions are found in certain parameter ranges~\cite{ref:33}. The question then arises whether in these phases the $\alpha$'s also form a condensate. In the past, we calculated critical temperatures for $\alpha$ particle condensation in infinite symmetric nuclear  matter studying a generalized Thouless criterion for quartets. Quite high values for $T_c^{\alpha}$ were obtained. However, the Thouless criterion applies for weak coupling situations and not for the strong coupling case with well established bosonic clusters. Therefore, the high $T_c^{\alpha}$ values obtained in Ref.~\cite{ref:34} are questionable. 

If we treat the $\alpha$'s as ideal elementary bosons, one obtains the following ideal Bose gas relation for $T_c^{\alpha}$, 
\begin{equation}
T_c^{\alpha}/{\rm MeV} = 13.634 (n_N/{\rm fm}^{-3})^{2/3},
\end{equation}
\noindent
where $n_N$ denotes the nucleon density. Thus, $T_c^{\alpha}$ = 1 MeV occurs at the nucleon density $n_N$ = 0.02 fm$^{-3}$. However, at that density interactions like the Pauli blocking become of importance, and the ideal Bose gas is no longer an adequate approximation.

For densities in the range (1/1000 - 1/10)$\rho_0$, one obtains low values for $T_c^{\alpha}$. However, this also is not a realistic estimate. What is needed is an interpolation formula covering for quartet condensation the density dependence of $T_c^{\alpha}$ from high (weak coupling) to very low and vanishing densities. Only in the last situation the above formula is strictly valid. In the case of pairing such an interpolation scheme between strong and weak coupling has been proposed by Nozi\`eres and Schmitt-Rink~\cite{ref:35}. It would be very important to generalize this scheme to the quartet case. Only then we can have a rough idea in which parameter range $\alpha$ particle condensation can occur in compact stars, i.e. proto neutron stars, etc. In this context another interesting problem to solve is the transition, i.e. disappearance of $\alpha$ particles with 
increasing density (Mott transition) and appearance of ordinary superfluidity (Cooper pairing in the deuteron channel). The latter being more stable at high density~\cite{ref:34,ref:36}. 

The formation of $\alpha$ matter and $\alpha$ particle condensation in infinite matter is an important issue in deriving the nuclear matter equation of state (EOS) at subsaturation densities. In the low-temperature region, below the Mott density the $\alpha$ particles yield the dominating contribution to the composition of symmetric matter, if we restrict us to clusters with $A \le 4$. Below densities of the order 10$^{-4}$ fm$^{-3}$, nuclear matter can be considered as an ideal mixture of free nucleons and clusters. The interaction between the constituents can be neglected.

The mass action law gives increasing yield of $\alpha $ particles at decreasing temperature for fixed density, but decreasing bound state concentration for decreasing density at fixed temperature (entropy dissociation). The low-density limit at fixed temperature is the ideal classical gas of nucleons. Corrections are given by the virial expansion. In particular, within a cluster-virial expansion the empirical scattering phase shifts can be used to evaluate density corrections for the ideal mixture of the different components. Thus, the virial coefficient for $\alpha-\alpha$ interaction is obtained from the corresponding scattering phase shifts \cite{horowitz/schwenk}.

Alternatively, one can use the phase shifts to introduce an effective interaction. The corresponding EOS has been reconsidered recently \cite{misicu}, also taking into account the formation of a quantum condensate. The suppression of the $\alpha$ condensate with increasing density was recently considered by Funaki et al.~\cite{funaki_2008} in context with the formation of a condensate-like state in low-density nuclei and its disappearance with increasing density. 

However, such effective interactions become questionable if the density is increasing so that the wave functions of the clusters overlap. Then, Pauli blocking leads to the dissolution of the clusters. Instead of an effective $\alpha$ particle interaction fitted to the scattering phase shifts, one has to go back to the constituent nucleon-nucleon interaction. The antisymmetrization of the wave function and the account of the Pauli exclusion principle become essential. Calculations on that level have been performed only for a lattice of $\alpha$ particles~\cite{akaishi69,brink73,tohsaki89,tohsaki96,takemoto}.  

The full calculation for $\alpha$ matter (freely moving  $\alpha$ particles) would give the energy per nucleon for symmetric matter in the low-density limit. Furthermore, it would be of importance to calculate the $\alpha$-pair distribution function in infinite matter, and to calculate the modification of the $\alpha$ form factor at increasing nucleon density within such a first principle approach that includes the full antisymmetrization of the nucleon wave function. The relevance for finite dilute nuclei and the outer region of heavy nuclei is obvious.

An important quantity is the binding energy per nucleon or, more generally, the nuclear matter EOS. At very low densities, the internal energy per nucleon takes the value $U=\frac{3}{2} T$ for the classical ideal gas of free nucleons. When clusters are formed in nucleon gas with low temparature and very low density, the binding energy of the nucleons in clusters determines the internal energy. In the case of $\alpha$ particles this contribution to the internal energy amounts to about $-7$ MeV. It determines the internal energy at zero temperature in the low-density limit. Recently~\cite{ref:33} appropriate approaches to the EOS have been investigated, to be of use in supernova explosion calculations.

With increasing density, the Pauli blocking leads to a reduction of the binding energy of the $\alpha$ particles and its dissolution. This is the reason for the increase of internal energy until the Mott density ($\approx$ 0.01 fm$^{-3}$) is reached. Above the Mott density, after the bound states are dissolved, the free nucleon Relativistic Mean field approach or similar effective quasiparticle approaches give the behavior of the internal energy. A more detailed investigation of the internal energy should also include the virial coefficient for $\alpha-\alpha$ interaction \cite{horowitz/schwenk,misicu} as well as the formation of a condensate \cite{takemoto} what has not been considered up to now in the EOS at low temperatures and low densities.

The approximation of the uncorrelated medium can be improved considering the cluster mean-field approximation~\cite{cmf,RSM,ropke:2008qk}. This would also improve the correct inclusion of $\alpha$ matter as discussed here. The formation of quantum condensates (quartetting) and its disappearance with increasing density demands further investigations. 

We  considered only the formation of light clusters $A \le 4$. This limits the parameter range of the total neutron density $n_{n}^{\rm tot}$, the total proton density $n_{p}^{\rm  tot}$, and $T$ in the phase diagram to that area where the abundances of heavier clusters are small. For a more general approach to the EOS which takes also the contribution of heavier cluster into account, see Refs.~\cite{RSM,hempel09}. Future work on the nuclear matter EOS will include the contribution of the heavier clusters.  

Phase separation occurs when thermodynamic stability $\partial \mu_N/\partial n_N \ge 0$ is violated. For given $T$, volume $\Omega$ and particle numbers $N_{\tau} = n_{\tau} \Omega$, the minimum of the free energy $F = {\cal{F}} \Omega$ has to be found. If the EOS $n_{\tau}=n_{\tau}(T,\mu_n, \mu_p)$ is known, this thermodynamic potential follows from integration, e.g., $F (T,n_{p},n_{n}) = \int_{0}^{n_{n}} \mu_{n}(T,0,n'_n) dn'_n +\int_0^{n_p} \mu_p(T,n'_p,n_n) dn'_p$. For stability against phase separation, the curvature matrix ${\cal{F}}_{\tau,\tau'}= \partial^{2}{\cal{F}}/\partial n_\tau\partial n_{\tau'}|_T$ has to be positive, i.e. Tr $[{\cal{F}}_{\tau,\tau'}] \ge 0$, Det $[{\cal{F}}_{\tau,\tau'}] \ge 0$.

At present, it is unclear whether the chemical potential becomes larger with increasing nucleon density near the region, where the bound states are dissolved, indicating a region of metastability. This would be of high interest for the metastability of low-density nuclei. The existence of a condensate will influence not only the EOS of nuclear matter, but also the transport properties and reaction rates (e.g. neutrino absorption) in such systems.

\section{$\alpha$ particle vs $^8$Be condensation}

In Bose-Einstein condensation for bosons with attractive interaction the question was discussed~\cite{ref:48} whether the bosons condense as singles or as molecules if the interaction is strong enough to bind two bosons to a molecule in free space. Without Coulomb repulsion $^8$Be would be bound by 3-5 MeV. In compact stars Coulomb is screened by electrons. Even in nuclei within a gas of $n\alpha$'s the Coulomb repulsion between two $\alpha$'s may be screened to some extent. It is intriguing that some studies of the Hoyle state show~\cite{ref:15_feldmeier, ref:16_enyo} that in the loosely bound $\alpha$ gas state always two of the three $\alpha$'s are slightly closer to one another than to the third $\alpha$ particle. So the question certainly is relevant whether one should consider the condensation of $\alpha$'s or the one of $^8$Be's. Of course, depending on temperature and/or asymmetry, also heavier bosonic clusters as $^{16}$O, $\cdots$, $^{56}$Fe can be formed. However, for the moment we do not consider this possibility, since the time scale to form these heavier clusters may be considerably longer than for the lighter ones. On the other hand, depending on temperature and/or asymmetry they may not be formed at all, see Sec.~\ref{sub:5}. Since $^8$Be is a loosely bound dimer of two $\alpha$'s, it may be appropriate to consider both clusters together.

\section{Gas of trimer clusters}

In nuclear physics we not only have clusters of even number of fermions (bosons) but also clusters of odd number of fermions (which again are fermions). For example nucleons are in first approximation strongly bound clusters of three quarks. Then at higher densities and/or temperatures it would be very interesting to know how these clusters dissolve into their constituents, or the other way round, a process known as hadronization. It is experimentally known that $^6$He has an excited state at 12-15  MeV where the $^6$He is broken up into two tritons. A gas of $n\alpha$ particles where $2n$ neutrons are added may have a resonant state of $n$ tritons. Formally such a gas of trimers again could be treated with a THSR wave function~\cite{ref:50_yokohama},

\begin{equation}
\Psi_{\rm trimer} = {\cal A}[\phi_t \phi_t \cdots \phi_t].
\end{equation}

\noindent
The antisymmetrizer will take care of the fact that the new gas of trimers builds up its proper Fermi surface. 
In the nucleon, two of the three quarks may form a strongly bound di-quark which can be considered as a boson. 
How boson-fermion correlations are built up in a boson-fermion mixture has 
 recently been treated in Refs.~\cite{watanabe08,barillier08}.

\section{Conclusions}

The field of $\alpha$ particle condensation is rapidly evolving. 
However, as discussed in this contribution, many open questions subsist. 
One of the most urgent issues is to verify experimentally that in $^{16}$O indeed the 15.1 MeV state
 is an $\alpha$ particle condensate state. 
It is very intriguing to explore Hoyle-analogue states in $A\not=4n$ nuclei. 
A correct determination of the critical temperature for $\alpha$ particle condensation in infinite matter
 is also very important in the astrophysical context.
On the other hand, the formation of cluster condensation is discussed in cold atom physics, 
 as experimenters are able to trap simultaneously more than two kinds of fermions. 
For example, one has succeeded to fabricate trions~\cite{ref:1} with three different fermions
 and in the future eventually also quartets with four different fermions. 
Whether quartets exist in the form of bi-excitons in semi conductors is still an open question~\cite{bi-exciton}. 
Nuclear physics, thus, is standing at the forefront of this subject.



\section*{References}

\begin{thebibliography}{10}

\bibitem{ref:2}
Y. Funaki, H. Horiuchi, W. von Oertzen, G. R\"opke, P. Schuck, A. Tohsaki, and T. Yamada, Phys. Rev. C. {\bf 80}, 064326 (2009).

\bibitem{thsr}
A.~Tohsaki, H.~Horiuchi, P.~Schuck, and G.~R\"opke, Phys. Rev. Lett. {\bf 87}, 192501 (2001).

\bibitem{tohsaki_nara}
A. Tohsaki, H. Horiuchi, P. Schuck, and G. R\"opke, Nucl. Phys. A {\bf 738}, 259 (2004).

\bibitem{yamada04}
T. Yamada and P. Schuck, Phys. Rev. C {\bf 69}, 024309 (2004).

\bibitem{funaki08}
Y. Funaki, T. Yamada, H. Horiuchi, G. R\"opke, P. Schuck, and A. Tohsaki, Phys. Rev. Lett. {\bf 101}, 082502 (2008).


\bibitem{monopole}
T. Yamada, Y. Funaki, H. Horiuchi, K. Ikeda, and A. Tohsaki, Prog. Theor. Phys. {\bf 120}, 1139 (2008).

\bibitem{ref:3}
B. W. Lynch, unpublished.

\bibitem{ref:4_borderie}
B. Borderie and M. F. Rivet {\it et al.}, to be published.

\bibitem{ref:5_chevalier}
P. Chevalier, F. Scheibling, G. Goldring, I. Plesser, and M. W. Sachs, 
Phys. Rev. {\bf 160}, 827 (1967).

\bibitem{ref:6_freer}
M. Freer {\it et al.}, Phys. Rev. C {\bf 51}, 1682 (1995); Phys. Rev. C {\bf 70}, 064311 (2004).

\bibitem{ikeda68}
K. Ikeda, N. Takigawa, and H. Horiuchi, Prog. Theor. Phys. Suppl. extra number, 464 (1968).

\bibitem{ref:7_tanihata}
P. Zarubin, private communication.


\bibitem{ref:9}
N. P. Andreeva et al., Eur. Phys. J. A {\bf 27}, 295 (2006).
 
\bibitem{ref:10_oertzen}
W. von Oertzen, Eur. Phys. J. A {\bf 29}, 133 (2006).

\bibitem{ref:11_brenner}
M. W. Brenner {\it et al.}, in Proceedings of the Int. Conf. ``Clustering 
Phenomena in Nuclear Physics'', St. Petersburg, published in Physics of 
Atomic Nuclei (2000, Yadernaya Fizika).

\bibitem{ref:12_ogloblin}
A. A. Ogloblin {\it et al.}, in Proceedings of the Int. Nuclear Physics Conf., Peterhof, Russia, June 28-July 2, 2005.

\bibitem{morinaga56}
H. Morinaga, Phys. Rev. {\bf 101}, 254 (1956); Phys. Lett. {\bf 21}, 78 (1966). 

\bibitem{ref:14} 
For example, Y.~Suzuki, H.~Horiuchi, and K.~Ikeda, Prog.~Theor.~Phys. {\bf 47}, 1517 (1972);
 B.~Fulton, Proc. 7th Int. Conf. on Clustering Aspects of Nuclear Structure and Dynamics, Rab, Croatia,
 1999, (World Scientific), pp.~122-129.

\bibitem{ref:15_feldmeier} 
M. Chernykh, H. Feldmeier, T. Neff, P. von Neumann-Cosel, and A. Richter, Phys. Rev. Lett. {\bf 98}, 032501 (2007).

\bibitem{ref:16_enyo} 
Y. Kanada-En'yo, Phys. Rev. Lett. {\bf 81}, 5291 (1998).

\bibitem{funaki_yamada09}
Y.~Funaki and T.~Yamada, private communication (2009).

\bibitem{ref:17_ohkubo_to_be_published} 
S. Ohkubo and Y. Hirabayashi, Phys.~Lett.~B {\bf 684}, 127 (2010).

\bibitem{ref:18_chappell} 
S. P. G. Chappell {\it et al.}, Phys. Rev. C. {\bf 51}, 51 (1995).

\bibitem{ref:19} 
M. Girardeau, J. Math. Phys. (N.Y.) {\bf 1}, 516 (1960); M. D. Girardeau, Phys. Rev. {\bf 139}, B500 (1965).

\bibitem{pitaevskii03}
L.~Pitaevskii and S.~Stringari, {\it Bose-Einstein Condensation}, Oxford Science Publications, International Series of Monographs on Physics, 116 (2003).

\bibitem{ref:20_itagaki} 
N. Itagaki, S. Okabe, K. Ikeda, and I. Tanihata, Phys. Rev. C {\bf 64}, 014301 (2001).

\bibitem{ref:21} 
D. H. Wilkinson, Nucl. Phys. A {\bf 452}, 296 (1986).

\bibitem{girod} 
M. Girod et al., to be published.

\bibitem{yamada08} 
T.~Yamada and Y.~Funaki, Int.~J.~Mod.~Phys.~E {\bf 17}, 2101 (2008).

\bibitem{sasamoto06} 
T. Kawabata et al., Int.~J.~Mod.~Phys.~E {\bf 17}, 2071 (2008).

\bibitem{yamada06} 
T.~Yamada, H.~Horiuchi, and P.~Schuck, Mod.~Phys.~Lett.~A {\bf 21}, 2373 (2006).

\bibitem{yamada09} 
T.~Yamada, talk at the Sixth Workshop on 
Aspects of alpha correlations and alpha condensation in nuclear systems,
Institute of Physics, Rostock University, Germany, 5th-7th August, 2009.

\bibitem{soic04}
N.~Soi\'c {\it et al.}, Nucl. Phys. A {\bf 742}, 271 (2004).

\bibitem{curtis05}
N.~Curtis {\it et al.}, Phys. Rev. C {\bf 72}, 044320 (2005).

\bibitem{charity08}
R.~J.~Charity {\it et al.}, Phys. Rev. C {\bf 78}, 054307 (2008).

\bibitem{lattimer_svesty}
  J.~M.~Lattimer and F.~D.~Swesty,
  Nucl.\ Phys.\  A {\bf 535}, 331 (1991).


\bibitem{toki-shen}
  H.~Shen, H.~Toki, K.~Oyamatsu, and K.~Sumiyoshi,
  Progr. Theor. Phys. {\bf 100}, 1013 (1998); 
  Nucl.\ Phys.\  A {\bf 637}, 435 (1998).


\bibitem{shimiyoshi-roepke}
K. Sumiyoshi and G. R\"opke, Phys. Rev. C {\bf 77}, 055804 (2008).


\bibitem{ref:33}
S.~Typel, G.~R\"opke, T.~Kl\"ahn, D.~Blaschke, and H.~H.~Wolter,
  Phys.\ Rev.\ C {\bf 81}, 015803 (2010).
 

\bibitem{ref:34}
G. R\"opke, A. Schnell, P. Schuck, and P.Nozi\`eres, 
Phys. Rev. Lett. {\bf 80}, 3177 (1998).

\bibitem{ref:35}
P. Nozi\`eres and S. Schmitt-Rink, J. Low Temp. Phys. {\bf 59}, 195 (1985).

\bibitem{ref:36}
T. Sogo, R. Lazauskas, G. R\"opke, and P. Schuck, Phys. Rev. C {\bf 79}, 051301 (2009).

\bibitem{horowitz/schwenk}
 C.~J.~Horowitz and A.~Schwenk,
  Nucl.\ Phys.\  A {\bf 776}, 55 (2006).

\bibitem{misicu}
G. R\"opke, Phys. Rev. C {\bf 79}, 014002 (2009).

\bibitem{funaki_2008}
Y. Funaki, H. Horiuchi, G. R\"opke, P. Schuck, A. Tohsaki, and T. Yamada, 
Phys. Rev. C {\bf 77}, 064312 (2008).

\bibitem{akaishi69}
Y.~Akaishi and H.~Band\=o, Prog. Theor. Phys. {\bf 41}, 1594 (1969).

\bibitem{brink73}
D. M. Brink and J. J. Castro, Nucl. Phys. A {\bf 216}, 109 (1973).

\bibitem{tohsaki89}
A. Tohsaki, Prog. Theor. Phys. {\bf 81}, 370 (1989); {\bf 88}, 1119 (1992); {\bf 90}, 871 (1993).

\bibitem{tohsaki96}
A. Tohsaki, Phys. Rev. Lett. {\bf 76}, 3518 (1996).

\bibitem{takemoto}
H.~Takemoto, M.~Fukushima, S.~Chiba, H.~Horiuchi, Y.~Akaishi, A.~Tohsaki, Phys. Rev. C  {\bf 69}, 035802 (2004).



\bibitem{cmf}
G. R\"opke, T. Seifert, H. Stolz, and R. Zimmermann, Phys. Stat. Sol. (b) {\bf 100}, 215 (1980).

\bibitem{RSM}
 G. R\"opke, M. Schmidt, L. M\"unchow, and H. Schulz,
 Nucl.\ Phys.\ A {\bf 379}, 536 (1982); 
 {\bf 399}, 587 (1983); {\bf 424}, 594 (1984).


\bibitem{ropke:2008qk}
G. R\"opke, Phys. Rev. C {\bf 79}, 014002 (2009).

\bibitem{hempel09}
M. Hempel and J. Schaffner-Bielich, arXiv:0911.4073 [nucl-th].

\bibitem{ref:48}
P. Nozi\`eres and D. Saint James, J. Physique {\bf 43}, 1133 (1982).

\bibitem{ref:50_yokohama}
P. Schuck, T. Sogo, and G. R\"opke, Int. J. Mod. Phys. A {\bf 24}, 2027 (2009).

\bibitem{watanabe08}
T. Watanabe, T. Suzuki, and P. Schuck, Phys. Rev. A {\bf 78}, 033601 (2008).

\bibitem{barillier08}
X. Barillier-Pertuisel, S. Pittel, L. Pollet, and P. Schuck, 
Phys. Rev. A {\bf 77}, 012115 (2008).

\bibitem{ref:1}
A. N. Wenz, T. Lompe, T. B. Ottenstein, F. Serwane, G. Zuern, and S. Jochim, 
Phys. Rev. A {\bf 80}, 040702(R) (2009).

\bibitem{bi-exciton}
S. A. Moskalenko and D. W. Snoke, Bose-Einstein Condensation of
Excitons and Bi-Excitons, Cambridge University Press, 2000.


\end{thebibliography}
\end{document}